# Stochastic syncing in sinusoidally driven atomic orbital memory

Werner M. J. van Weerdenburg[1,†], Hermann Osterhage[1,†], Ruben Christianen[1], Kira Junghans[1], Eduardo Domínguez[2], Hilbert J. Kappen[2], Alexander Ako Khajetoorians[1,*]

[1]*Institute for Molecules and Materials, Radboud University, 6525 AJ Nijmegen, the Netherlands*

[2] *Donders Institute for Neuroscience, Radboud University, 6525 AJ Nijmegen, the Netherlands*

*corresponding author: a.khajetoorians@science.ru.nl

[†]both authors contributed equally

**Stochastically fluctuating multi-well systems as physical implementations of energy-based machine learning models promise a route towards neuromorphic hardware. Understanding the response of multi-well systems to dynamic input signals is crucial in this regard. Here, we investigate the stochastic response of binary orbital memory states derived from individual Fe and Co atoms on a black phosphorus surface to sinusoidal input voltages. Using scanning tunneling microscopy, we quantify the state residence times for DC and AC voltage drive with various input frequencies. We find that Fe and Co atoms both exhibit features of synchronization to the AC input, but only Fe atoms demonstrate a significant frequency-dependent change in the time-averaged state occupations. By modeling the underlying stochastic process, we show that the frequency response of the system is directly related to the DC voltage dependence of the state asymmetry. This relation provides a tunable way to induce population changes in stochastic systems and lays the foundation for understanding the response of multi-well systems to dynamical input signals.**

**Introduction**

The development of neuromorphic hardware that is inherently adaptive and exhibits self-learning requires a deeper understanding of materials that show multi-well behavior (*1*). One route toward this end is to create materials systems with tunable multi-well energy landscapes, where local minima represent information (*2*). Such systems are mimicked by energy-based models in machine learning, for instance the Hopfield model or its stochastic analogue, the Boltzmann machine, where the energy landscape can be described by an Ising Hamiltonian (*3, 4*). In these models, each spin represents an individual neuron and the coupling is represented by a synaptic, memory-bearing weight (*5*). To this end, there have been a number of tunable coupled spin arrays derived from atomic spins on surfaces, based on scanning tunneling microscopy (*6-9*). While the coupling in such spin systems can be tuned, these platforms are often limited to double-well energy landscapes and thus to binary configurations (*10, 11*). However for interesting neuromorphic computing, it was proposed that atomic spin ensembles require competing long range interactions (*12*), linked to the concepts of spin glasses (*13-16*). Recently, it was shown that arrays of coupled binary orbital memory states emulate an atomic-scale Boltzmann machine, by exploiting their dynamical properties (*17, 18*). The multi-state dynamics emerging from the coupling between stochastic atoms directly maps onto the Hamiltonian of the Boltzmann machine, with tunable stochastic weights. A local gate voltage can be used to influence the system dynamics, and thus tune the stochastic weights. (*17, 19*).

Fundamentally, it is still unclear how such stochastic multi-well systems respond to a dynamical drive signal. For an individual binary spin in the stochastic regime, the response to a periodic stimulus is often described in terms of phase-synchronization and stochastic resonance (SR) (*20-22*). Generally, SR manifests itself in nonlinear dynamical systems as an enhanced response to a weak input signal due to the presence of intermediate noise, compared to the response for low and high noise amplitudes. In systems with an activation threshold energy, SR arises when a weak, noisy signal periodically exceeds the threshold, thereby inducing dynamics. This is referred to as threshold SR (*21*). However, in Arrhenius-type double-well systems where the switching rates are always finite, similar phenomena can be observed due to a periodic modulation of the switching rates, known as dynamical SR (*20, 21*). Recently, dynamical SR has also been reported in atomic-scale systems, where the dynamics of the system originates from quantum tunneling processes (*23, 24*). It remains both theoretically and experimentally unclear how multi-state systems, where multiple Markov rates are present, respond to a periodic drive (*25*). Experimentally, coupled arrays of orbital memory present a promising route toward exploring stochastic response in the multi-well limit. However, since the dynamical complexity rapidly increases with the number of stochastic units (*17*), it is essential to first gain a fundamental understanding of how synchronization and frequency-induced population coding can be achieved with AC signals for an individual atom.

Here, we quantify the stochastic response of orbital memory derived from individual Fe and Co atoms on black phosphorus (BP) to sinusoidal input voltages. Using scanning tunneling microscopy (STM), we apply a chosen DC bias voltage and AC sinusoidal voltage and study the stochastic response of the orbital states of individual Fe and Co atoms as a function of frequency and amplitude. For individual Fe atoms, we observe a frequency-dependent change in the state-dependent residence times. This modulation results from synchronization between the input signal and the switching events. Concomitantly, we find a change in the extracted state favorability, namely the asymmetry, near the frequency corresponding to the mean residence time of the two orbital states of Fe. In contrast, we do not observe a frequency-dependent change in the asymmetry for individual Co atoms, even though there is a significant synchronization response in the measured residence time statistics. The origin of this effect is related to the voltage-dependent switching rates of each system. To quantify this effect, we measure the change in asymmetry as a function of the DC bias voltage, or namely the voltage susceptibility, for both Fe and Co. We relate the synchronization behavior and frequency-dependent changes in the asymmetry to the voltage susceptibility in the DC limit. Based on a stochastic model, we demonstrate that these observations are dictated by the switching rates for each state, in the DC limit, for the given atom.

**Results**

We performed scanning tunneling microscopy and spectroscopy (STM/STS) on Fe and Co atoms adsorbed on a black phosphorus (BP) surface using a home-built UHV-STM system (*26*) ($p < 5 \cdot 10^{-10}$ mbar), operating at a temperature of $T \approx 7$ K. Individual Fe and Co atoms, residing in the hollow site of BP, exhibit orbital memory: a bistability in their valency (*18, 27*). The two stable valency configurations can be distinguished in constant-current STM images by both their apparent height ($\Delta z$), and the shape of their charge density (Fig. 1A,B). We label each of these two distinguishable states as high/low for each atomic species. Below a threshold DC voltage, each orbital state remains indefinitely stable allowing for state identification.

Above an applied DC voltage ($V_{DC}$), there is a finite probability of switching between the two orbital configurations for both individual Co and Fe atoms (*18, 27*). The stochastic switching is recorded by measuring the tunneling current ($I_t$) as a function of time for a constant tip-sample separation and lateral tip position. Before each measurement, the tip is first stabilized on the BP substrate to a pre-defined tip-sample separation. Fig. 1C exemplifies the stochastic switching for an individual Fe atom ($Fe_{low}$/$Fe_{high}$). By recording a statistical number of stochastic switching events (> 1000 events), we extract the average residence time ($\bar{\tau}_{low/high} = \frac{1}{N_{low/high}} \sum_i \tau_{i,low/high}$) of each orbital state, where $\tau_i$ refers to one measured residence time with index $i$ and $N_{low/high}$ is the total number of detected switching events for a given state (see Fig. S3 for details about the data analysis). Moreover, from the average residence times we extract the asymmetry $A = (\bar{\tau}_{low} - \bar{\tau}_{high})/(\bar{\tau}_{low} + \bar{\tau}_{high})$, which measures the state favorability. The values of $\bar{\tau}_{low/high}$ and $A$ are sensitive to $V_{DC}$, $I_t$ and the position of the tip. We selectively choose parameters to yield switching rates within the time resolution of the experiment (see Fig. S2) and avoid the strongly asymmetric regime for these experiments.

In addition to measuring the stochastic switching for a given $V_{DC}$, we also measured the stochastic switching in response to an additional sinusoidal voltage $V_{AC} = |V_{AC}|/2 \sin(2\pi f t)$ added to $V_{DC}$ (Fig. S2). $|V_{AC}|$ is subsequently defined as the peak-to-peak amplitude. The resultant values of $I_t$ are modulated, as shown in Fig. 1D, for the case where $|V_{AC}|$ = 40 mV and $f$ = 2 Hz. The number of switching events is highest when $|V_{DC} + V_{AC}|$ is maximized, indicating that the switching probability of each state $i$ is sensitive to the phase $\varphi$ of $V_{AC}$. Moreover, the switching probability remains finite at any point of the waveform. This means that we consider synchronization of a thermodynamic double-well system to a sinusoidal input signal (*20, 21*).

In Fig. 2A,B, we illustrate histograms of the state-resolved $\tau_i$ for a set of frequencies, for both atom types ($|V_{AC}|$ = 40 mV). Each data point represents the number of detected $\tau_i$'s within a given time interval. The histograms contain the exponential decay behavior, expected for a Poisson process. In addition to the exponential decay, above a given frequency, there are clear deviations from the pure exponential distribution and additional modulations emerge on top of the exponential decay. Such features are typical of phase-synchronization in a stochastic system, indicating that switching preferentially occurs around one well-defined phase $\varphi$. This results in a higher favorability of residence times with $\tau_i \sim n/f$, where $n$ is an integer. Such features are also present in systems that exhibit SR (20) and have been observed for atomic spins derived from Fe atoms (24). For individual Co atoms, we also observe synchronization for both states, albeit with less statistics, given the longer residence times of the Co system. For a given $|V_{AC}|$, the amplitude of the modulations depends on the specific state and atom type.

In order to further quantify the stochastic response, we extracted the values of $\bar{\tau}_{low}$, $\bar{\tau}_{high}$ and $A$ as a function of $f$ (Fig. 2C,D). For Fe, $A(f)$ illustrates a strong change near the frequency $f_{Fe}^* \approx 1/\bar{\tau}_{av}$, where $\bar{\tau}_{av}$ corresponds to the average residence time of both states at the applied value of $V_{DC}$. Above $f_{Fe}^*$, $\bar{\tau}_{low}$ and $\bar{\tau}_{high}$ increase and decrease, respectively, and saturate to asymptotic values. Since the change in $A(f)$ leads to a change in the time-averaged occupation and each orbital state is distinguished by a different tunneling current, the frequency response of individual Fe atoms can be detected in the time-averaged current signal (see Fig. S9). Therefore, the change in $A(f)$ represents the time-averaged frequency response of the system. In contrast to the significant frequency response of both $\bar{\tau}$ and $A$ for an Fe atom, there is no significant frequency response in the various extracted variables for a Co atom, as plotted in Fig. 2E,F. We explain this difference below.

In order to understand the difference in the frequency response between Fe and Co atoms, we first quantified $\bar{\tau}_{low/high}$ as a function of $V_{DC}$ without an additional $V_{AC}$ signal (Fig. 3A,B). The general $V_{DC}$–dependent trend for both Co and Fe agrees with previous publications (18, 27), where $\bar{\tau}_{low/high}$ decreases for an increasing magnitude of $|V_{DC}|$. Phenomenologically, we can describe $\bar{\tau}_{low/high}(V_{DC})$ in the voltage range of interest using an exponential dependence, as shown with the fits in Fig. 3A,D. In addition, we also extracted the dependence of $A$ on $V_{DC}$ and find that $A(V_{DC})$ in the studied range strongly depends on the value of $V_{DC}$ for the case of Fe, but is nearly independent of $V_{DC}$ for Co. Accordingly, we can relate the slope of $A$, which we define as $\chi = \partial A / \partial V_{DC}$. We refer to $\chi$ as the voltage susceptibility. From Fig. 3A,D, we observe that $|\chi_{Fe}| \gg |\chi_{Co}|$, which, as we show below, is directly linked to the frequency response. Besides $\bar{\tau}$ and $A$, the switching characteristics can be described in terms of the state occupations, defined as $n_{low} = \bar{\tau}_{low} / (\bar{\tau}_{low} + \bar{\tau}_{high})$ and $n_{high} = \bar{\tau}_{high} / (\bar{\tau}_{low} + \bar{\tau}_{high})$.

We subsequently model the stochastic response of each atom type to the applied $V_{AC}$. We assume that $\tau_i$ in absence of $V_{AC}$ is described by a Poisson process with an exponential probability distribution. For $V_{AC} = 0$, the distribution of $\tau_i$ can be described by average transition rates $w_{low \to high}(V_{DC}) = 1/\bar{\tau}_{low}(V_{DC})$ and $w_{high \to low}(V_{DC}) = 1/\bar{\tau}_{high}(V_{DC})$, which are extracted from Fig. 3A,D. Subsequently, we simulate the stochastic dynamics in response to a non-zero value of $V_{AC}$, considering a Poisson point process with periodically changing $w$, as introduced in ref. (*28*). Since $A(f)$ is a time-averaged quantity, it can be simulated by solving the associated rate-equations (see Supplementary information). We find that the simulated $A(f)$ (solid line) is in good agreement with the experimental data in Fig. 2D,F. To model the $\tau_i$ histograms, stochastic data can be generated using the experimentally determined $w_{low \to high}(V_{DC})$ and $w_{high \to low}(V_{DC})$ with the aforementioned assumptions (see Materials and Methods). This approach allows us to model the $\tau_i$ histograms (Fig. 2A,B), as well as $n_{low}$ and $n_{high}$ as a function of the modulation phase $\varphi$.

Based on this modeling, we considered the average $n_{low}$, $n_{high}$, $w_{low \to high}$, and $w_{high \to low}$ within a period of the waveform. In Fig. 3B, we plot the phase-resolved $n_{low}(\varphi)$ (red) and $n_{high}(\varphi)$ (blue), for various frequencies. Each data point is extracted by subdividing one period of the waveform into 100 phase segments, calculating $n_{low}$, $n_{high}$, $w_{low \to high}$, and $w_{high \to low}$ for each segment and averaging over many periods. For $f \ll f^*$, $n_{low}$ and $n_{high}$ are clearly modulated, directly following the changes of $A(V_{DC})$ in Fig. 3A. For instance, at $\varphi = 3/2\,\pi$ and $f$ = 1 Hz, the Fe_high state is more favorable, partially compensating for the larger $n_{low}$ in the first half of the waveform. With increasing $f$, the modulation of $n_\text{low}$ and $n_\text{high}$ becomes less pronounced and the maximum of $n_\text{high}$ shifts to larger $\varphi$, indicating that $n_\text{low}$ and $n_\text{high}$ exhibit a delayed response in time to $V_\text{AC}$. For $f \gg f^*_\text{Fe}$, $n_\text{low}$ and $n_\text{high}$ show a flat response as a function of $\varphi$, converging to a time-averaged A. We find that the evolution of $n_\text{low}$ and $n_\text{high}$ as a function of $f$ is reproduced. The phase-resolved conditional transition rates, $w_\text{low→high}(\varphi)$ and $w_\text{high→low}(\varphi)$, can be determined from the switching rates within each phase segment, divided by the occupation of the corresponding state in that phase segment (*28*). As shown in Fig. 3C, most of the switches occur around $\varphi = \pi/2$, as expected from $\bar\tau_\text{low/high}$ in the voltage range of $V_\text{DC} \pm |V_\text{AC}|/2$, and $w_\text{low→high}(\varphi)$ and $w_\text{high→low}(\varphi)$ are approximately constant as a function of *f*. The phase-resolved analysis for individual Co atoms reveals that $w_\text{low→high}(\varphi)$ and $w_\text{high→low}(\varphi)$ (Fig. 3F) are clearly modulated, while $n_\text{low}$ and $n_\text{high}$ (Fig. 3E) shows negligible changes within one period of the waveform. This can be understood as a direct consequence of the parallel evolution of $\bar\tau_\text{low}$ and $\bar\tau_\text{high}$, namely $\chi_\text{Co} \approx 0$ for the two Co states in Fig. 3D. As a result, the frequency-induced change of $A(f)$ (Fig. 2E,F) is zero for Co atoms.

Based on the comparison between the dynamical properties of Fe and Co atoms and supported by stochastic modeling, we can identify the requirements to observe a time-averaged frequency response and link it to the measured $\bar\tau_\text{low/high}(V_\text{DC})$. Namely, the frequency-induced changes of $A(f)$ can only be observed if $n_\text{low}$ and $n_\text{high}$ are modulated for $f \ll f^*$. This can only be obtained when $w_\text{low→high}(V)$ and $w_\text{high→low}(V)$ evolve such that $A(V)$ exhibits significant changes in the voltage range of $V_\text{DC} \pm |V_\text{AC}|/2$ (i.e. $\chi$ must be finite). Furthermore, we can expect the change in $A(f)$ to occur for $f^* \approx 1/\bar\tau_\text{av}$.

To generalize the relation between $\bar\tau_\text{low/high}(V_\text{DC}|V_\text{AC} = 0)$ and the resulting frequency response, we studied the stochastic response of individual Fe atoms for different input signals. Firstly, we apply an input voltage with a fixed $|V_\text{AC}|$ = 40 mV around three different values of $V_\text{DC}$, as illustrated in Fig. 4A. Each of the cases shows a similarly sized change of $A(f)$ with the previously observed sigmoidal shape. This is expected, given the nearly constant value of $\chi_\text{Fe}$ in the voltage range of interest. However, the sign change of $A(V_\text{DC})$ in Fig. 4A introduces a vertical offset for the three cases.

Additionally, a shift of the $A(f)$ curves in Fig. 4B along the frequency axis with varying $V_{DC}$ can be seen. This reflects the lower $\bar{\tau}_{\text{low,high}}$ of the system at higher $V_{DC}$.

To further illustrate the tunability of the frequency response in the state occupation, we measured $A(f)$ for variable $|V_{AC}|$. We probed the frequency response for $|V_{AC}|$ = 20, 30, 40 and 50 mV, while oscillating around the same bias point of $V_{DC}$ = -510 mV, as indicated in Fig. 4C. We find that the total change of $A(f)$ keeps increasing as $|V_{AC}|$ increases (Fig. 4D). Interestingly, $A(f \ll f^*)$ saturates around 0 for each $|V_{AC}|$. These observations suggest that the system adopts an average $A$ with an effective bias $V_{\text{eff}} \approx V_{DC}$. The saturated $A(f \gg f^*)$ indicates that the application of $V_{AC}$ mimics the application of a $V_{DC}$, where $V_{\text{eff}}$ shifts towards shorter $\bar{\tau}_{\text{low/high}}$ with increasing $f$. The relation between the $\bar{\tau}_{\text{low/high}}(V_{DC})$ and the time-averaged frequency response was statistically confirmed by measuring the change of $A(f)$ for multiple Fe atoms using similar conditions (see Fig. S7). Additionally, we could obtain varying trends of $A(f)$ by modifying the rate evolution with the tip position, $V_{DC}$ and the atom type (see Fig. S8). Each of these data sets could be reproduced by simulating $A(f)$ based on a rate-equation model, as exemplified in Fig. 4B,D.

**Discussion**

In conclusion, we studied the stochastic response of the orbital memory states derived from individual Fe and Co atoms to a sinusoidal drive. For individual Fe and Co atoms, we observed phase-synchronization for residence times that coincide with the inverse of the applied drive frequencies. This is manifested by a deviation from a Poisson distribution, compared to the DC limit. Notably, this effect can be observed for frequencies much higher than the inverse average residence time. Based on stochastic modeling, we can relate the amplitude of the synchronization effect to the evolution of the switching rates as a function of bias voltage. As a second effect, we observed a change in the state favorability, namely the asymmetry, as a function of frequency. In contrast to the response of individual Fe atoms, this effect is absent for individual Co atoms. Based on modeling and measuring the state favorability and lifetimes as a function of applied DC voltage, in the absence of a sinusoidal drive, we found that both the synchronization effect and the frequency-dependent asymmetry are derived from the underlying voltage-dependent transition rates. This is an important consideration, when using a sinusoidal drive and the resultant change in asymmetry, as a measure of the mean residence time of a two-state system, when the telegraph processes cannot be resolved. In other words, a time-averaged frequency response can only be observed if the modulated switching rates exhibit a significant asymmetry change in the modulation range of interest. If no change in asymmetry with frequency is observed, the system may still show dynamics and synchronization in that frequency range. We showed that the asymmetry and residence times of single-atom orbital memory can be controlled by tuning the tip position, $V_{DC}$ and atom type. This in turn enables tuning the frequency response of orbital memory to an AC stimulus. The characteristic frequency is dictated by the transition rates and can be tuned by the injection current $I_t$ (27). The magnitude of the frequency response depends on $\chi$ in the modulation range and can be tuned by choice of the atom type, tip position, $V_{DC}$, and $V_{AC}$. This is advantageous when probing multi-well behavior, as for example seen in the atomic Boltzmann machine. As coupled orbital memory exhibit multi-well behavior (17), these studies provide a starting basis for future studies where the stochastic response of multi-state systems to complex waveforms can be interrogated. We expect coupling atoms that switch stochastically at different time scales to respond in multiple frequency regimes.

**Acknowledgments**

**Funding**

The experimental part of this project was supported by the European Research Council (ERC) under the European Union's Horizon 2020 research and innovation programme (grant no. 818399). This publication is part of the project 'What can we 'learn' with atoms?' (with project number VI.C.212.007) of the research programme VICI which is (partly) financed by the Dutch Research Council (NWO). We also acknowledge the support of the Interdisciplinary Research Platform of the Faculty of Science at the Radboud University.

**Author contributions**

WMJvW, HO, RC and KJ performed the experiments. HO and ED performed the model calculations. WMJvW, HO, RC, and KJ performed the data analysis, while all authors discussed and contributed to the design and interpretation of the data analysis. AAK designed the experiments. All authors contributed to the writing of the manuscript and scientific discussion. HJK and AAK supervised the project.

**Competing interests**

Authors declare that they have no competing interests.

**Data and materials availability:**

All data needed to evaluate the conclusions in the paper are present in the paper or the Supplementary Materials. Data for all figures presented in this study are available at Radboud Data Repository. The following software was used: Nanonis 4.5 and LabVIEW 2015 (experiments), Gwyddion 2.60 and Matlab R2019b (data processing), Jupyter 7.0 (data simulations), Adobe Illustrator 27.8.1 (figure preparation).

**Materials and Methods**

We performed scanning tunneling microscopy and spectroscopy (STM/STS) using a home-built UHV-STM system (*26*) ($p < 5·10^{-10}$ mbar), operating at a temperature of $T \approx 7$ K. All measurements were performed with electrochemically etched W tips. Prior to approaching black phosphorus (BP), these tips are dipped and characterized on a clean Au(111) surface. We prepare individual Fe and Co atoms adsorbed on black phosphorus (BP) by *in-situ* cleaving a bulk BP crystal (HQ graphene) and (co)-depositing Fe and/or Co while cooling the substrate to ~30 K.

To combine $V_{DC}$ and $V_{AC}$ signals, we use the setup illustrated in Fig. S2. The AC modulation is provided by an arbitrary waveform generator (Keysight 33600 A Series) and added to $V_{DC}$ using an active adder. The combined signal as well as the current signal are each recorded using a multifunction input/output device (NI USB-6251) for data acquisition.

**Statistical Analysis**

Switching events in the measured current trace were detected by defining a threshold current between two separated current levels (Fig. S3A) after subtraction of a linear background and after further subtraction of an oscillating current contribution in form of a sine wave for $V_{AC} \neq 0$. The residence times in between switching events $\tau_{i,\text{low/high}}$ and their statistical distribution were further used in the analysis. Fig. S3B exemplarily shows histograms of $\tau_{\text{low/high}}$ for $V_{DC}$ = -500 mV, $V_{AC}$ = 0 mV. Their distribution is described by an exponential function with decay constant $\frac{1}{\bar{\tau}_{\text{low/high}}}$, where $\bar{\tau}_{\text{low/high}}$ is the average residence time $\frac{1}{N_{\text{low/high}}}\sum_i \tau_{i,\text{low/high}}$. This exponential distribution is the characteristic of a Poisson point process.

To quantify uncertainties in the measurement of $\bar{\tau}$, we define a standard deviation $\sigma$ of $\bar{\tau}$ by averaging random sets of 500 $\tau_{i,\text{low/high}}$ acquired under equivalent experimental conditions and calculating the standard deviation of these averaged values. Error bars throughout the paper represent $1\sigma$.

**Simulating the state occupations for AC voltage input**

The transition rates in the orbital memory are strongly bias dependent (*18,27*). Accordingly, $V_{\text{AC}}$ modulates the transition rates in the bistable orbital memory periodically. In order to simulate the state occupation of the orbital memory in the AC regime, we developed a model of a Poisson point process with transition rates changing periodically in time. The stepwise approach of simulating the system dynamics and time-averaged state occupations from this stochastic model will be described in the following.

First, we derive the voltage dependence of the transition rates $w_{i \to j}$ between the binary orbital memory states. Fig. 4A exemplarily shows the mean state lifetimes of a single Fe atom for varying DC bias $V_{\text{DC}}$. The transition rates are derived from the mean value of the measured state lifetimes: $w_{i \to j} = \frac{1}{\bar{\tau}_i}$, ($i$ = high/low). We perform a weighed exponential fit in order to derive a functional dependence of the transition rates, on the applied bias voltage (see Figure S4A):

$$w_{i \to j}(V) = a \cdot \exp(b|V|).$$

In the fitting procedure, the values of $w_{i \to j}$ for each bias value are weighed with the inverse of their squared standard deviation $\frac{1}{\sigma^2}$. The choice of an exponential function is purely phenomenological. We constrain the fit on the bias interval [$V_{\text{DC}}$ - $V_{\text{AC}}$/2, $V_{\text{DC}}$ + $V_{\text{AC}}$/2] that is relevant for simulating the AC response. Following the exponential bias dependence, we model the time dependent transition rates like

$$w_{i \to j}(t) = a \cdot \exp\left(b\left[V_{DC} + \frac{1}{2}V_{AC} \sin(2\pi f t)\right]\right).$$

The time evolution of the state occupation $n_i(t)$ is then calculated numerically, starting from an equal distribution of both states $\left(n_{\text{low}}(0) = n_{\text{high}}(0) = \frac{1}{2}\right)$:

$$n_i(t + \Delta t) = n_i(t) + \Delta t \frac{dn_i}{dt},$$

with

$$\frac{dn_i}{dt} = n_j(t)w_{j \to i}(t) - n_i(t)w_{i \to j}(t).$$

Fig. S4B shows the time evolution of the state occupations obtained with this model.

**Simulations of stochastic real-time data for AC bias input**

The DC residence time statistics are accurately described by exponential distributions, corresponding to homogeneous Poisson point processes. The switching dynamics with fixed rates (constant voltage) can therefore be simulated as a repetition of two consecutive random steps, where exponentially distributed waiting times are generated with the state-dependent rates.

The non-homogeneous case, corresponding to a finite $V_{AC}$, is in principle not much more complex. A Poisson process with a time-dependent rate can be simulated by noticing that the integrated rate $I(t) = \int_0^t w(\tau) d\tau$ is a random variable that is exponentially distributed according to $P(I) = \exp(-I)$. The idea is then to generate a random value for $I$ and invert $I(t)$ to get a random sample of the residence time. As explained in the previous section, the switching rates $w(t)$ are fitted to an exponential of the voltage signal. For the AC case, the integrated rate function is inverted numerically via a lookup table. The fact that the rate is a periodic function of time is exploited for efficiency.

The synthetic stochastic data reproduces the experimental data for both Fe and Co atoms, as detailed in Figs. S5 and S6.

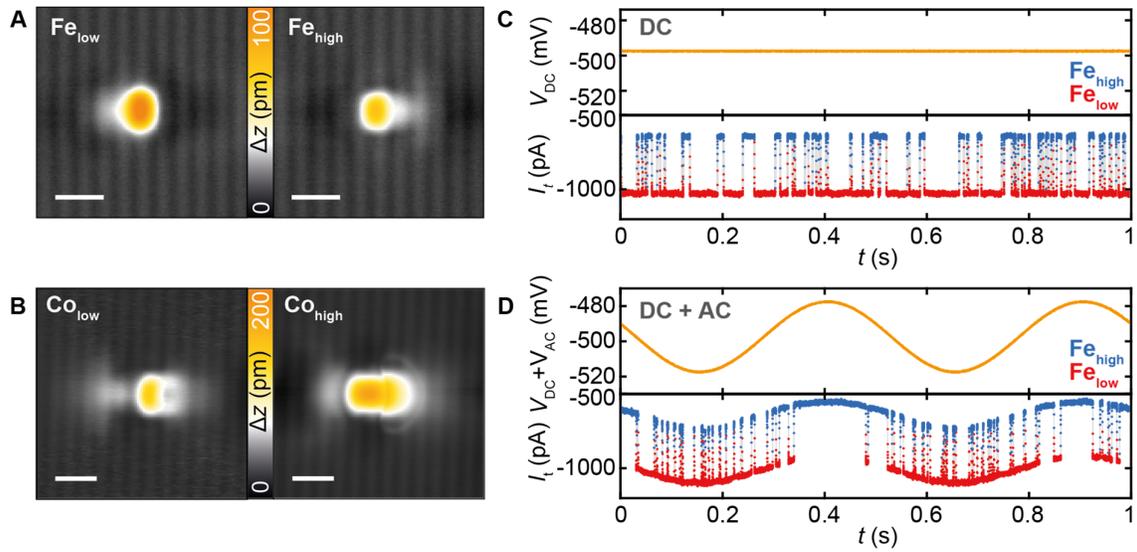

**Fig. 1. Stochastically switching orbital memory states of Fe and Co atoms on black phosphorus.** (A,B) Constant-current STM images of (A) an individual Fe atom and (B) an individual Co atom adsorbed in the hollow site of BP, imaged for both orbital memory states (low/high) ($V_{DC}$ = -400 mV, (A) $I_t$ = 10 pA and (B) 50 pA, scale bar = 1 nm). (C,D) Telegraph noise recorded at the center of an Fe atom in constant height for: (C) $V_{AC}$ = 0 mV, $V_{DC}$ = -500 mV and (D) $V_{AC}$ = 40 mV, $V_{DC}$ = -500 mV, $f$ = 2 Hz. The tip was stabilized on the substrate at $V_{DC}$ = -400 mV and $I_t$ = 100 pA, before the feedback loop was opened.

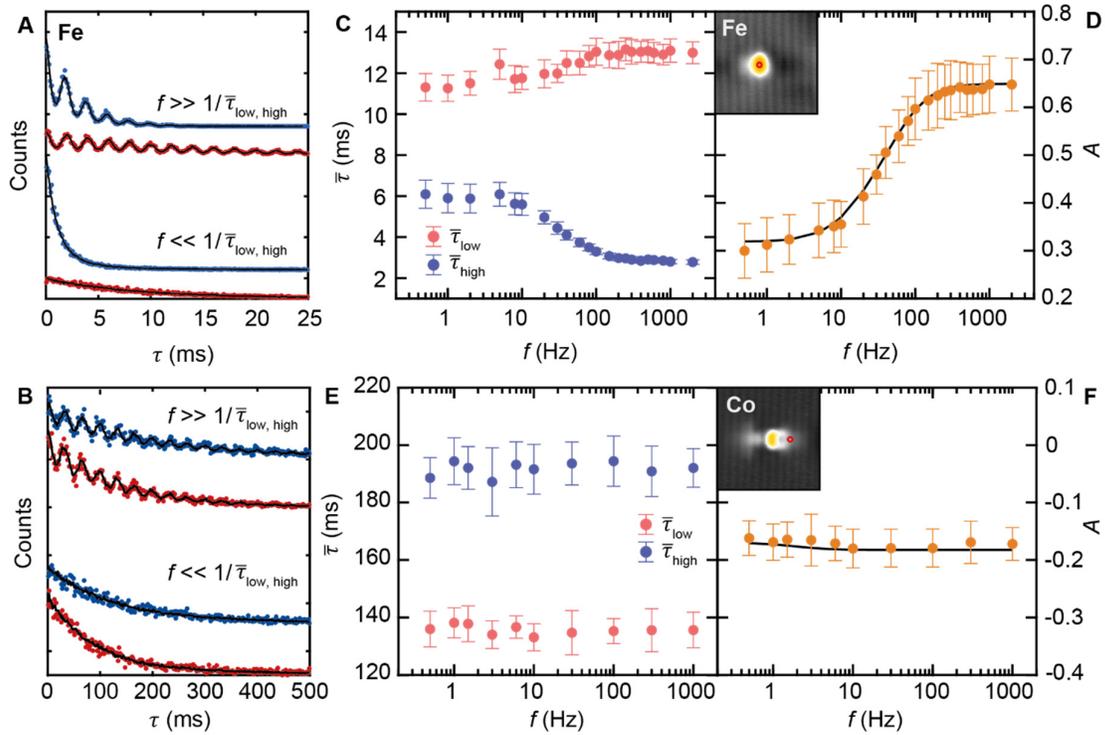

**Fig. 2. Frequency dependence of mean residence times.** (A,B) Histogram plots of the detected $\tau_i$ for (A) Fe$_{low}$ (red) and Fe$_{high}$ (blue), and (B) Co$_{low}$ (red) and Co$_{high}$ (blue) for two frequencies, measured with (A) $V_{DC}$ = -500 mV and (B) $V_{DC}$ = 480 mV with an amplitude of |$V_{AC}$| = 40 mV (peak-to-peak). The histograms present ≥ 28 extracted residence times per bin, i.e. per data point (plotted with an artificial vertical offset for each state and frequency). Histograms of synthesized stochastic data based on a Poisson point process are shown in black. (C,D) Time-averaged residence time $\bar{\tau}_{low/high}$ and deduced state asymmetry $A$ of Fe$_{low}$ (red) and Fe$_{high}$ (blue) as a function of the frequency of $V_{AC}$. (E,F) Time-averaged residence time $\bar{\tau}_{low/high}$ and deduced state asymmetry $A$ of Co$_{low}$ (red) and Co$_{high}$ (blue) as a function of the frequency of $V_{AC}$. Solid lines in (D,F) represent the simulated frequency response, based on rate-equations (see Materials and Methods). The tip was stabilized on the substrate at $V_{DC}$ = -400 mV and (A,C,D) $I_t$ = 100 pA or (B,E,F) $I_t$ = 30 pA.

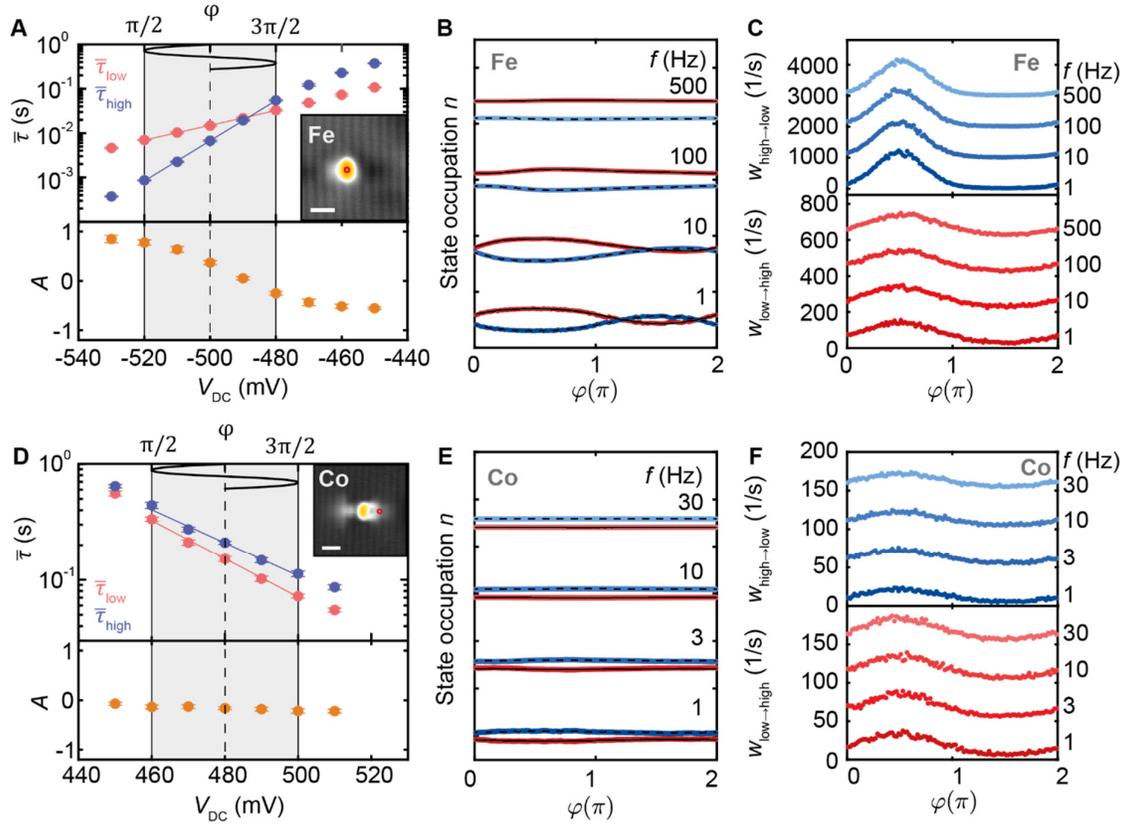

**Fig. 3. Voltage dependence of state asymmetry.** (A) Average residence times $\bar{\tau}_{low/high}$ for Fe$_{low}$ (red) and Fe$_{high}$ (blue) and extracted asymmetry $A$ (lower panel) as a function of bias voltage $V_{DC}$, measured with $V_{AC}$ = 0 mV on top of an individual Fe atom (see inset). (B) Phase-resolved state occupation $n_{low}$ (simulation: solid black line) and $n_{high}$ (simulation: broken black line) and (C) conditional switching rates $w_{low \rightarrow high}$ and $w_{high \rightarrow low}$ for different frequencies (artificially offset), averaged over many periods. $|V_{AC}|$ is indicated by the gray shaded area and the phase φ is defined at the top of (A). Tip height was stabilized on the substrate at $V_{DC}$ = -400 mV and $I_t$ = 100 pA. (D) Average residence times $\bar{\tau}_{low/high}$ for Co$_{low}$ (red) and Co$_{high}$ (blue) and extracted asymmetry $A$ (lower panel) as a function of bias voltage $V_{DC}$, measured with $V_{AC}$ = 0 mV next to an individual Co atom (see inset). (E) Phase-resolved state occupation $n_{low}$ (simulation: solid black line) and $n_{high}$ (simulation: solid black line) and (F) conditional switching rates $w_{low \rightarrow high}$ and $w_{high \rightarrow low}$ for different frequencies (artificially offset), averaged over many periods. $|V_{AC}|$ is indicated by the gray shaded area and the phase φ is defined at the top of (D). Tip height was stabilized on the substrate at $V_{DC}$ = -400 mV and $I_t$ = 30 pA.

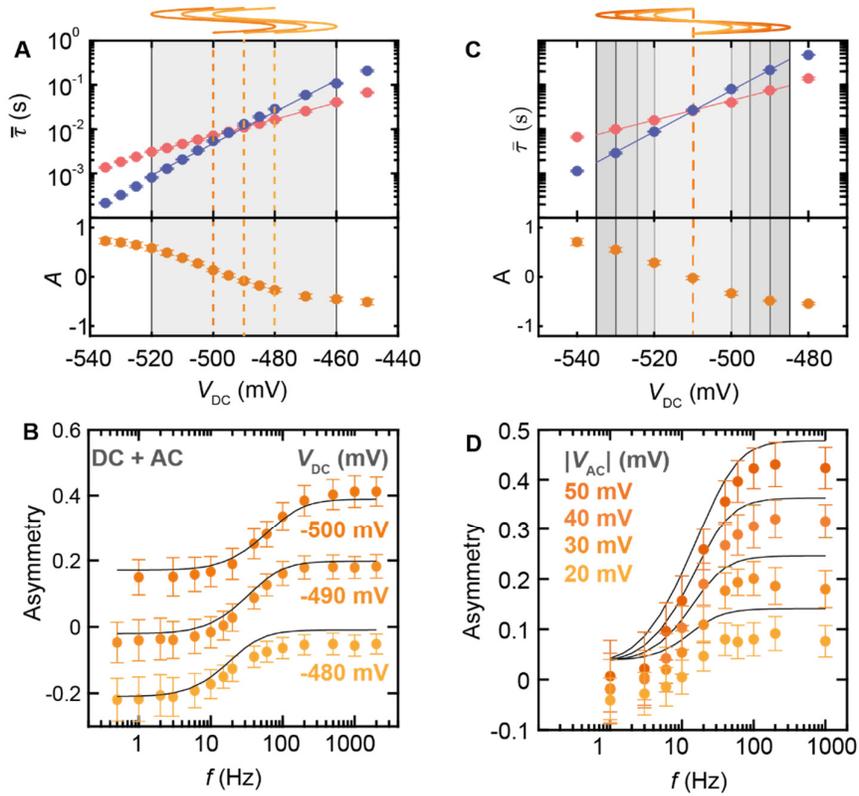

**Fig. 4. Frequency response in dependence of DC offset and AC amplitude.** (A) Average residence times $\bar{\tau}_{low/high}$ for Fe$_{low}$ (red) and Fe$_{high}$ (blue), and extracted asymmetry (lower panel) as a function of bias voltage $V_{DC}$, with $|V_{AC}|$ = 0 mV. (B) Average state asymmetry $A$ as a function of frequency with a constant $|V_{AC}|$ = 40 mV and three different $V_{DC}$ values, as indicated in (A) with the yellow lines. (C) Average residence times $\bar{\tau}_{low/high}$ for Fe$_{low}$ (red) and Fe$_{high}$ (blue), and extracted asymmetry (lower panel) as a function of bias voltage $V_{DC}$, measured with $V_{AC}$ = 0 mV. (D) Average state asymmetry $A$ as a function of frequency, measured with a constant $|V_{DC}|$ = -510 mV and four different $|V_{AC}|$ values, as indicated in (C) with the gray shaded areas. Solid lines in (B) and (D) represent the simulated frequency response based on the rate-equation model and the exponential fits (solid lines in (A) and (C)). Tip height was stabilized on the substrate at $V_{DC}$ = -400 mV and $I_t$ = 100 pA.

# Supplementary Materials for

## Stochastic syncing in sinusoidally driven atomic orbital memory

Werner M. J. van Weerdenburg[1,†], Hermann Osterhage[1,†], Ruben Christianen[1], Kira Junghans[1], Eduardo Domínguez[2], Hilbert J. Kappen[2], Alexander Ako Khajetoorians[1,*]

[1]*Institute for Molecules and Materials, Radboud University, 6525 AJ Nijmegen, the Netherlands*

[2] *Donders Institute for Neuroscience, Radboud University, 6525 AJ Nijmegen, the Netherlands*

*corresponding author: a.khajetoorians@science.ru.nl

[†]both authors contributed equally

## S1. Simulations of state occupation for AC bias input

Fig. S4B shows the result of a simulation for an Fe atom on BP under experimental conditions as in Fig. 2A of the main paper with an applied AC frequency of $f$ = 10 Hz. It can be seen that the expectation value of $n_i$ quickly converges to a steady state where it oscillates in time with frequency $f$. In order to deduce the time averaged state asymmetry $A = n_{\text{low}} - n_{\text{high}}$ shown in Fig. 2D, we calculate the average state occupations (broken lines) within two full periods of the AC signal in the equilibrium region (shaded area).

## S2. Simulations of stochastic real-time data for AC bias input

Fig. S5 and S6 show that the synthetic stochastic data (black solid lines) reproduces the synchronization observed for both, the Fe and the Co atom (Fig.S5A, S6A), as well as the different frequency response in the phase-resolved state occupations (Fig. S5B, S6B). Furthermore, it can be seen that the simulation, which is solely based on DC switching rates, reproduces the experimentally measured phase-resolved switching rates very well, underlining the feasibility of the description as a Poisson process with varying transition rates.

## S3. Data statistics

The tunability of the switching rates of the orbital memory states can provide a lot of variability to the frequency response of the system, as exemplified in Fig. 4. In Fig. S7, we present additional data sets on different Fe atoms, taken with identical measurement conditions as Fig. 2 and 3. We note that these measurements were not taken with an identical tip apex. In each case, small variations in the DC switching rates (A-C) impact the frequency dependent residence times (D-F), but the overall trend is consistent between all data sets.

Further variation was found by using different values of $V_{\text{DC}}$ and $I_t$, as shown in the additional data sets in Fig. S8. Moreover, Fig. S8F demonstrates that an individual Co atom can exhibit a (small) frequency response when $\chi_{\text{Co}}$ is finite (compared to Fig. 2 and 3 in the manuscript).

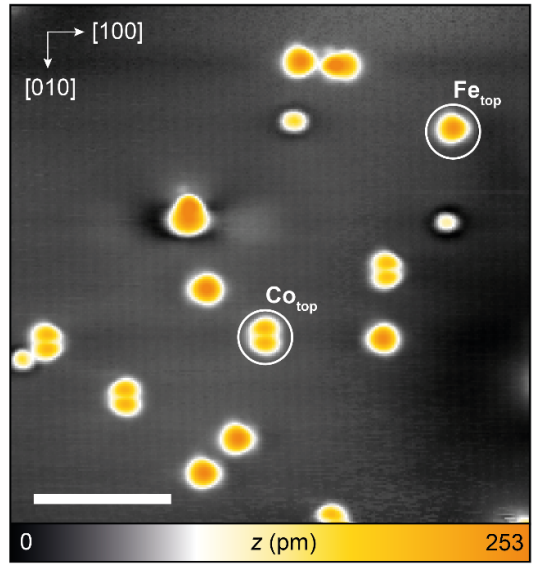

**Figure S1:** Constant-current STM image of the surface of black phosphorus with Co and Fe atoms adsorbed on the top binding site after deposition. To obtain well-isolated Fe and Co atoms, atoms can be laterally manipulated along the [010] direction and contaminated species (18) as well as BP vacancies (29) are avoided ($V_{DC}$ = -400 mV, $I_t$ = 10 pA, scale bar = 5 nm).

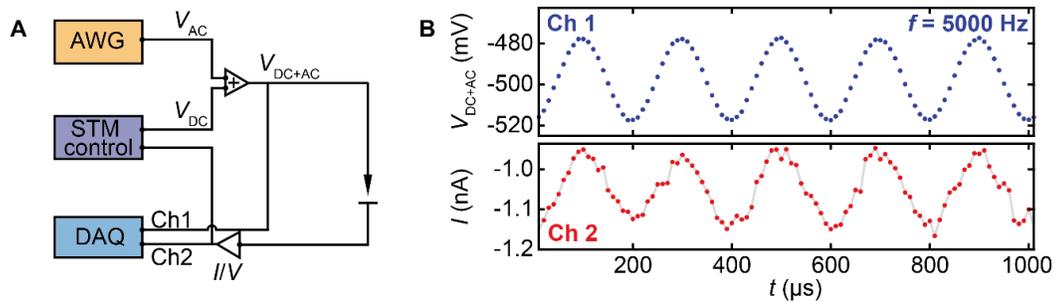

**Figure S2:** (A) Sketch of the experimental setup to measure with and without $V_{AC}$. (B) Example of a high-resolution time trace of $V_{DC+AC}$ on Channel 1 with a frequency $f$ = 5000 Hz and $I_t$ on Channel 2, recorded with a sampling rate of 100000 samples/s for each channel. Note that the AC modulation has the correct amplitude ($|V_{AC}|$ = 40 mV) and that no significant phase delay is detected between the two channels at $f$ = 5000 Hz.

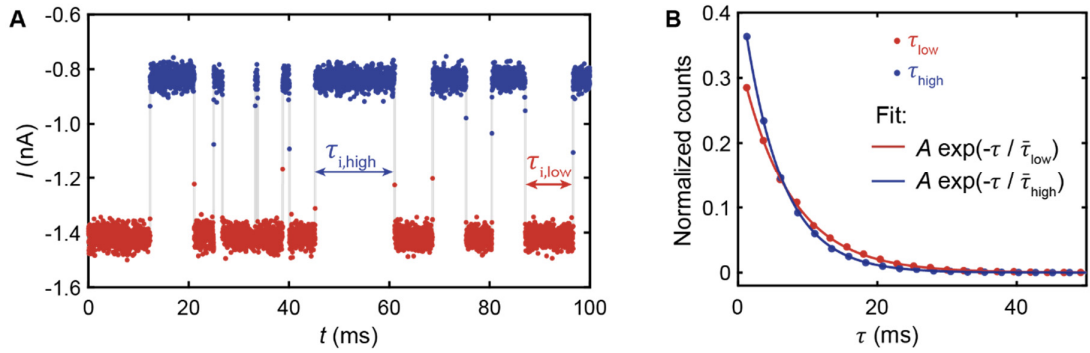

**Figure S3:** (A) Current trace of stochastic switching of a single Fe atom on BP measured with $V_{DC}$ = -500 mV ($V_{AC}$ = 0 mV). Single residence times $\tau_{i,\text{low/high}}$ for the low and high state are exemplarily indicated. The tip height was stabilized on the substrate at $V_{DC}$ = -400 mV and $I_t$ = 100 pA. (B) Residence time histogram for Fe$_{\text{low}}$ (red) and Fe$_{\text{high}}$ (blue). The plots present a total of 28589 residence times, separated into 25 bins per state. Solid lines represent exponential functions with $\bar{\tau}$, where amplitude parameter $A$ is fitted.

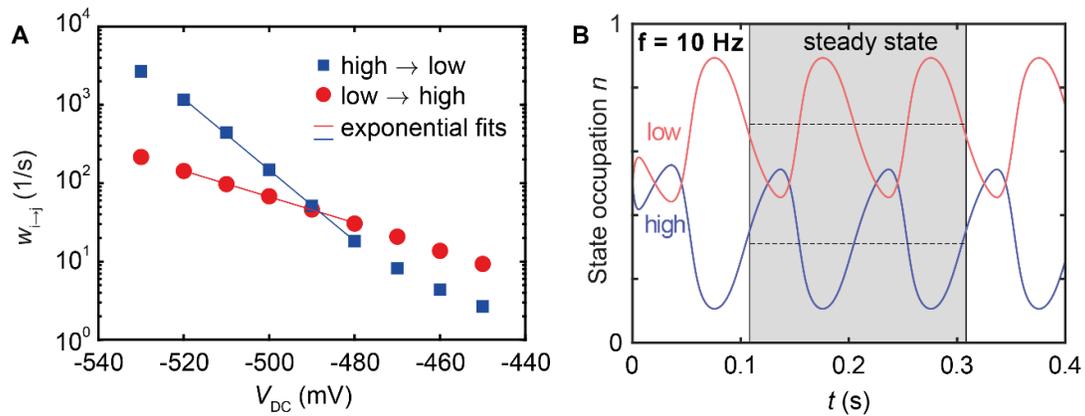

**Figure S4:** (A) DC transition rates $w_{i \to j}$ with exponential fits and (B) simulated state occupation as a function of time for an Fe atom on BP, according to the experimental conditions in Fig. 2C: $V_{DC}$ = -500 mV, $|V_{AC}|$ = 40 mV (peak-to-peak), tip stabilized on the substrate at $V_{DC}$ = -400 mV and $I_t$ = 100 pA. Here, the state occupation in the case of $f$ = 10 Hz is exemplarily shown. The simulation was initialized at $t$ = 0 in $n_{\text{low}} = n_{\text{high}} = \frac{1}{2}$. The time-averaged occupations (broken lines) in equilibrium conditions are averaged over exactly two periods of the AC signal (shaded area).

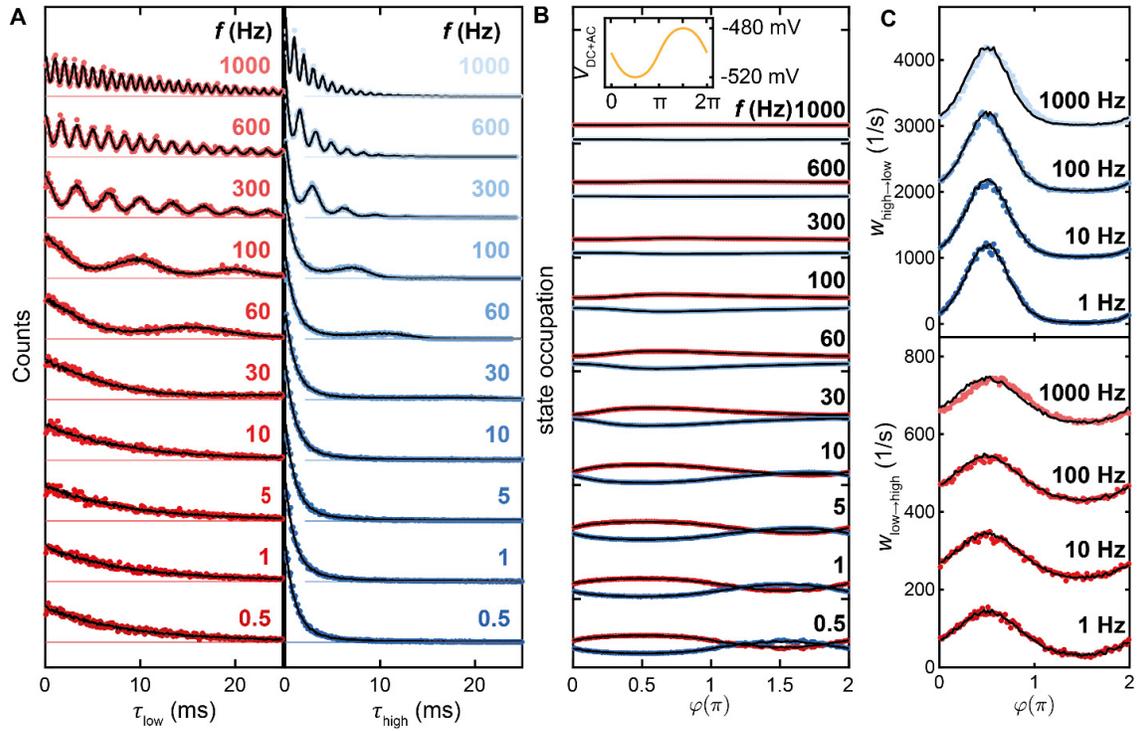

**Figure S5:** (A) Histograms of the residence times of $Fe_{low}$ (red) and $Fe_{high}$ (blue) for different frequencies (artificially offset). (B) Phase-resolved state occupation and (C) switching rate for different frequencies (artificially offset), averaged over many periods. The inset in (B) shows one period of the oscillatory bias voltage and defines the phase φ on the *x*-axis. Time-averaged properties of this data set are presented in Fig. 2D. The tip height was stabilized on the substrate at $V_{DC}$ = -400 mV and $I_t$ = 100 pA. Solid lines represent the results of stochastic simulations based on the DC switching rates $w_{i \to j}(V)$.

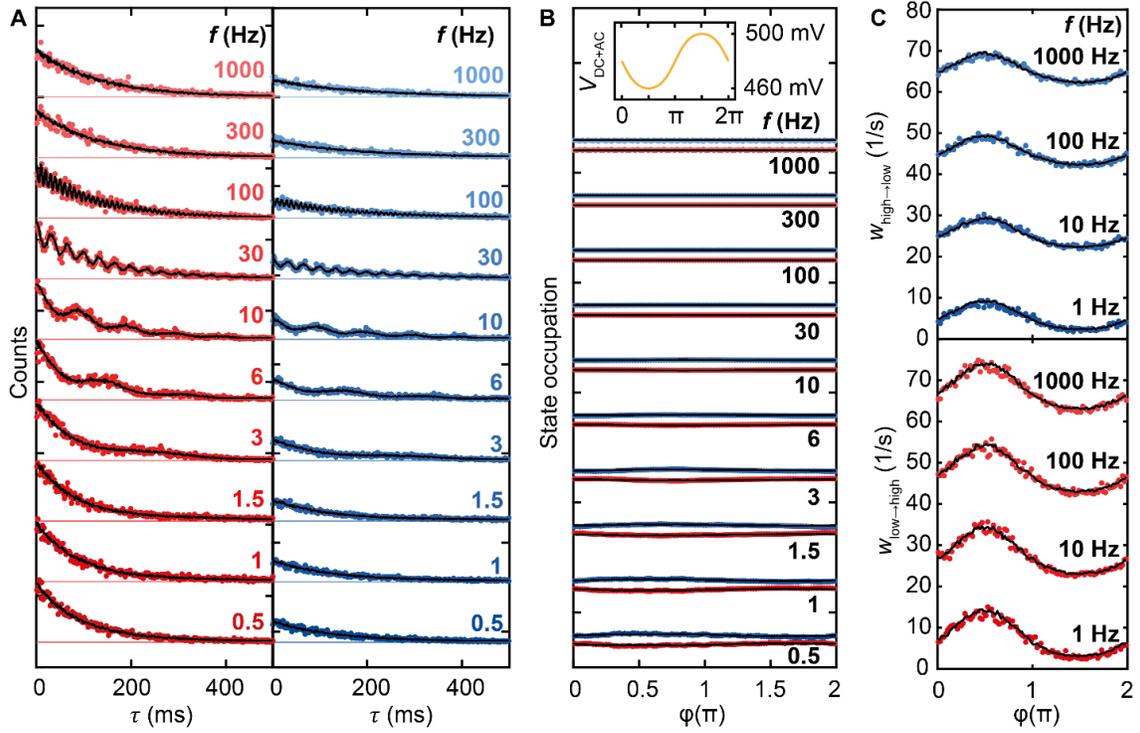

**Figure S6:** (A) Histograms of the residence times of $Co_{low}$ (red) and $Co_{high}$ (blue) for different frequencies (artificially offset). (B) Phase-resolved state occupation and (C) switching rate for different frequencies (artificially offset), averaged over many periods. The inset in (B) shows one period of the oscillatory bias voltage and defines the phase φ on the *x*-axis. Time-averaged properties of this data set are presented in Fig. 2D. The tip height was stabilized on the substrate at $V_{DC}$ = -400 mV and $I_t$ = 30 pA. Solid lines represent the results of stochastic simulations based on the DC switching rates $w_{i \to j}(V)$.

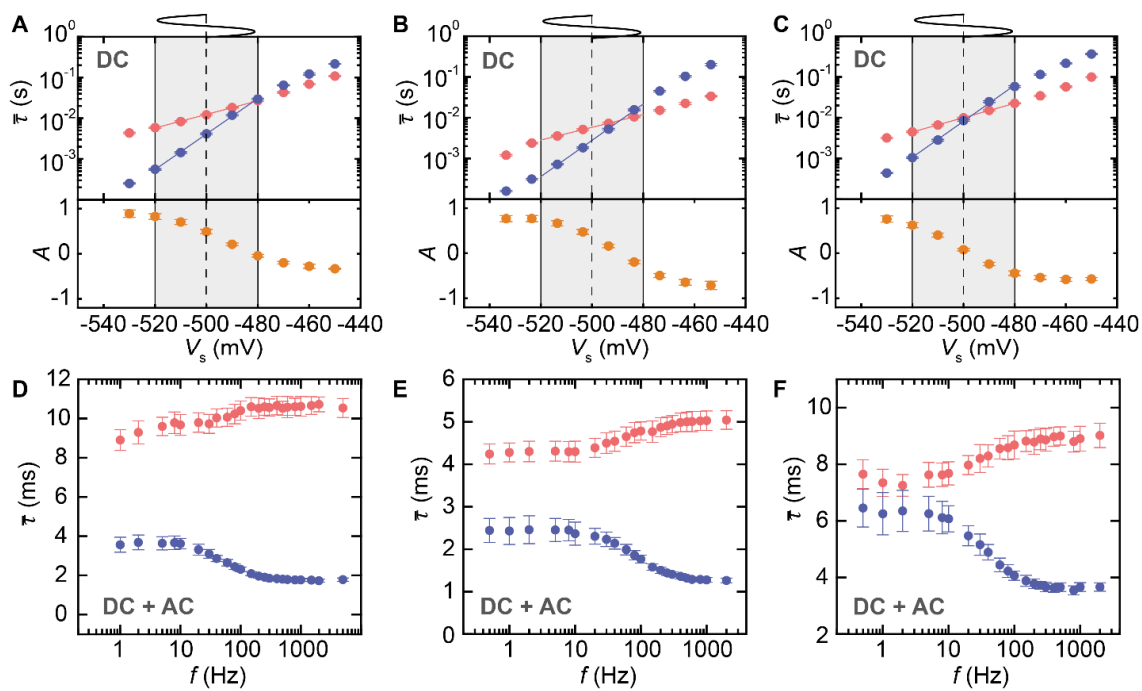

**Figure S7:** (A-C) Average residence times and extracted asymmetry, measured with $V_{AC}$ = 0 mV for different Fe atoms and tips, and (D-F) the corresponding frequency response of the average residence times. The data sets are taken with measurement conditions comparable to the data set presented in Fig. 2B; each data set was measured with $|V_{AC}|$ = 40 mV and $V_{DC}$ = -500 mV, and the tip height was stabilized on the substrate at $V_{DC}$ = -400 mV and $I_t$ = 100 pA.

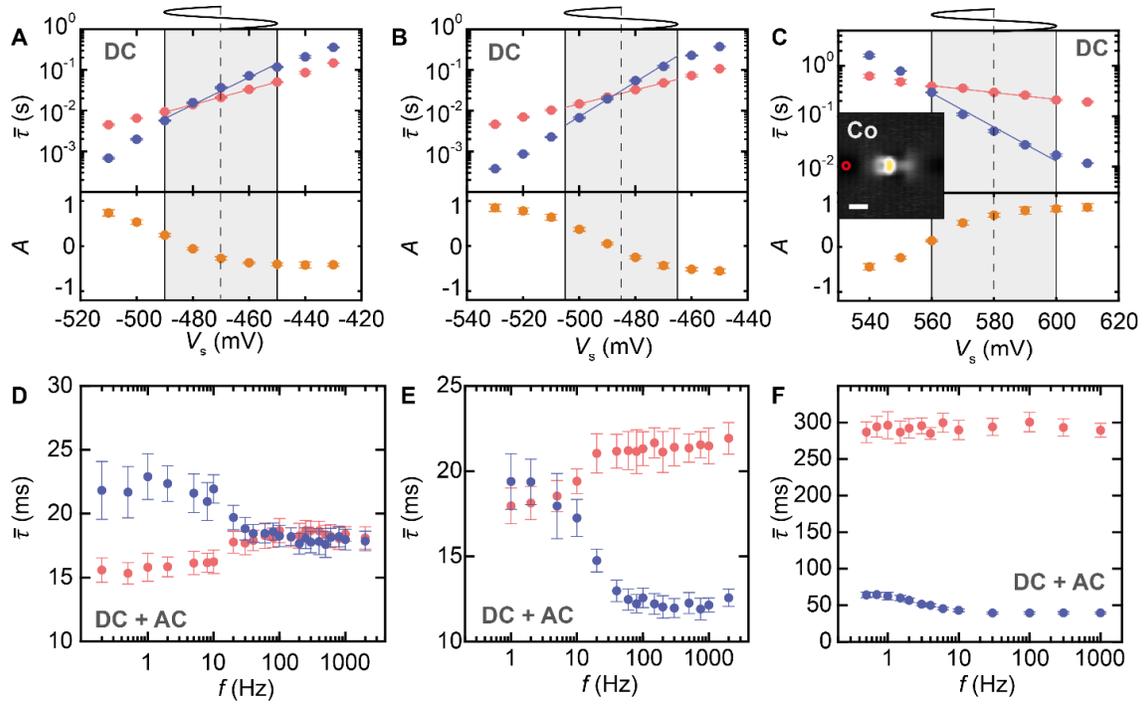

**Figure S8:** (A-C) Average residence times and extracted asymmetry ($V_{AC}$ = 0 mV) and (D-F) the corresponding frequency response of the average residence times. These data sets were measured (A,D) on top of an individual Fe atom using $V_{DC}$ = -470 mV, (B,E) on top of an individual Fe atom using $V_{DC}$ = -485 mV, and (C,F) next to an individual Co atom using $V_{DC}$ = 580 mV. Each data set was measured with $|V_{AC}|$ = 40 mV and with a different tip and atom. The tip height was stabilized on the substrate at $V_{DC}$ = -400 mV and (A,D) $I_t$ = 130 pA, (B,E) $I_t$ = 100 pA (C,F) $I_t$ = 70 pA.

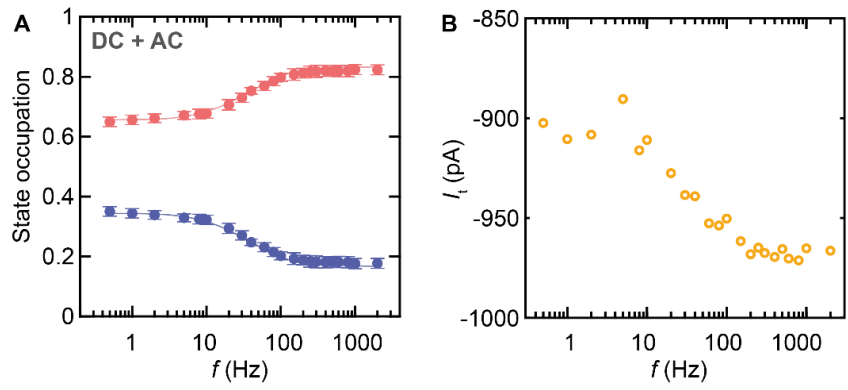

**Figure S9:** Comparison between the frequency-induced change in (A) the state occupation and (B) the time-averaged current. Data set corresponds to the data presented in Fig. 2(A,C,D) and Fig. 3(A-C).